\documentclass[11pt,twoside]{article}

\usepackage{asp2014}

\aspSuppressVolSlug
\resetcounters

\begin{document}

\title{Software metadata: How much is enough?}

\author{Alice Allen,$^{1,2}$ Peter Teuben,$^2$ G. Bruce Berriman,$^3$ Kimberly DuPrie,$^{4,1}$ Keith Shortridge,$^5$ and Rein Warmels$^6$}

\affil{$^1$Astrophysics Source Code Library, College Park, MD, US; \email{aallen@ascl.net}}
\affil{$^2$Astronomy Department, University of Maryland, College Park, MD, US}
\affil{$^3$Caltech/IPAC-NExScI, Pasadena, CA, US}
\affil{$^4$Space Telescope Science Institute, Baltimore, MD, US}
\affil{$^5$Knave and Varlet, McMahons Point, NSW, Australia}
\affil{$^6$European Southern Observatory, Garching, Germany}

% This section is for ADS Processing.  There must be one line per author.
\paperauthor{Alice Allen}{aallen@ascl.net}{}{Astrophysics Source Code Library}{}{College Park}{MD}{}{US}
\paperauthor{Peter Teuben}{teuben@astro.umd.edu}{orcid.org/0000-0003-1774-3436}{University of Maryland}{Astronomy Department}{College Park}{MD}{}{US}
\paperauthor{G. Bruce Berriman}{gbb@ipac.caltech.edu}{orcid.org/0000-0001-8388-534X}{California Institute of Technology}{IPAC-NExScI}{Pasadena}{CA}{}{US}
\paperauthor{Kimberly DuPrie}{kduprie@stsci.edu}{}{Space Telescope Science Institute/ASCL}{}{Baltimore}{MD}{}{US}
\paperauthor{Keith Shortridge}{keithshortridge@gmail.com}{orcid.org/0000-0003-1480-9217}{Knave and Varlet}{}{McMahons Point}{NSW}{}{Australia}
\paperauthor{Rein Warmels}{rwarmels@eso.org}{0000-0002-9814-0305}{European Southern Observatory}{}{Garching}{}{}{Germany}

%\ssindex{software}
%\ssindex{metadata}
%\ssindex{ASCL}

\begin{abstract}
Broad efforts are underway to capture metadata about research software and retain it across services; notable in this regard is the CodeMeta project. What metadata are important to have about (research) software? What metadata are useful for searching for codes? What would you like to learn about astronomy software? This BoF sought to gather information on metadata most desired by researchers and users of astro software and others interested in registering, indexing, capturing, and doing research on this software. Information from this BoF could conceivably result in changes to the Astrophysics Source Code Library (ASCL) or other resources for the benefit of the community or provide input into other projects concerned with software metadata.
\end{abstract}

%\ssindex{ASCL}
%\ssindex{software}
%\ssindex{metadata}

\section{Introduction}
That software is used in astronomy research is not news, and questions arise as how best to store, cite, index, discover, and maintain this software, and what information about it to track. These questions are common across all disciplines; the CodeMeta project\footnote{\url{https://codemeta.github.io/}} is a cross-disciplinary effort that seeks to provide a way to retain software metadata as it moves across services \citep{codemeta}, for example, from Astrophysics Source Code Library (ASCL) to NASA's Astrophysics Data Service (ADS) or GitHub to Zenodo\footnote{Respectively, {\url{https://ascl.net}}, {\url{http://adsabs.harvard.edu/}}, {\url{https://github.com/}}, {\url{https://zenodo.org/}}}, by creating and maintaining a crosswalk table for these metadata and the services that use them.

This BoF focuses on what metadata are most useful to astronomers under different circumstances, and is the latest in a series of BoFs that the ASCL has organized at ADASS focused on the needs of software authors and users. Previous sessions dealt with issues around sharing, reusing, crediting, and citing software; though these earlier BoFs touched upon software metadata, this is the first session to have these metadata as a central theme.

\section{Opening presentation and breakout groups}
Allen opened the session with information on broad efforts across many disciplines to define necessary metadata for software under different use cases. Metadata users typically want as much metadata as they can get, however, the more maintenance of metadata one has to do, the less likely it is that this information will be kept up to date. The ASCL, for example, has kept the amount of metadata it contains to a minimum, as previous other similar efforts, such as the Astronomical Software Directory Service (ASDS) \citep{asds}, folded in part because of the difficulties of maintaining metadata sufficiently to keep the resource useful \citep{allenschmidt}. Allen noted that the ASCL has often been asked by developers, journals, and users to include much more metadata and that the Journal of Open Source Software (JOSS) \citep{joss}, started in 2016, has also adopted a minimal metadata set very similar to that maintained by the ASCL.

She requested that participants break into groups to discuss and define the eight most important metadata elements for four different roles involved in the creation and use of this information. Eight was an admittedly arbitrary number, as any number would be, and was to try to enforce a hierarchy among all possible metadata that might be captured.

Participants were free to choose one of the following groups:

\begin{itemize}
\item Software Developer, moderated by Peter Teuben
\item Software User, moderated by Keith Shortridge
\item Software Indexer, moderated by Kimberly DuPrie
\item Journals, moderated by Alice Allen
\end{itemize}

The groups met for about 30 minutes and then came back together to share what they had discussed. 

\section{Results of breakout discussions}
Three of the breakout groups started with essentially a clean slate, discussing what the most important/basic information is of interest for the purposes of the role they were discussing. In contrast to this, the Software Indexer group started with a base set of metadata already defined \textemdash that currently captured by the ASCL \textemdash and discussed what information might be added to this set. Each moderator reported their results, consisting of a list of eight (or more) elements that would be of primary interest to the audience they were representing. These data were captured in a Google document\footnote{\url{http://tinyurl.com/BoFSoftwareMetadata}} that is open for viewing and commenting. 

Each of the groups had to determine what use cases they were trying to satisfy. The Journals group spent some time discussing the different use cases journals may have for metadata, including metadata needed to describe methods in the literature, and that for citation or software provenance. Without trying to describe formal use cases, the Users group mainly envisaged situations where someone had located an apparently attractive code, but wanted to know how easy it would be for them to make use of it in practice. They considered metadata that would describe support for the code (\textit{e.g.}, when last updated, name of current maintainer, if any) and ease of installation (\textit{e.g.}, system requirements, dependencies, packaging). The `family history' of a code was thought to be of interest\textemdash is this code a descendant of some other code, or conversely, is there another code available that is based on this? Although hard to express in metadata, many codes can be described as ``like some other package, but able to do such and such'', and all these things are useful to know. There was also discussion in the Users group about comments and ratings, potentially useful, but clearly controversial, and this group also raised the question ``Who supplies the metadata?''.

The Software Indexers group drifted into discussion of metadata that reflected the desires of software authors and users; this is understandable, as the group was composed almost entirely of people who write and use software rather than index it. The information this group shared reinforced the usefulness of metadata elements identified in other groups while also providing unique properties, such as whether a project had been forked. The Software Developers group, the largest of the four groups, identified the requested eight metadata properties, and went on to identify eight additional elements. This group provided a number of elements, such as \textit{preferred citation}, \textit{test results}, and \textit{API}, that were unique in this effort.

It is not surprising there were elements in common identified by the groups (see Table \ref{tab:common} for examples), and also not surprising there were many that were unique to one group (examples are shown in Table \ref{tab:unique}). This clearly demonstrates the differing needs of the various roles and the difficulty in getting the right amount and the right kind of metadata for the various users of it. What metadata one needs or would like to have is dependent on how that information will be used. The CodeMeta schema has 56 metadata properties in common with Schema.org's, and an additional ten properties to help satisfy the needs of all the use cases identified by that project. 

\begin{table}[]
\centering
\caption{Sample of metadata elements identified by more than one group}
\label{tab:common}
\begin{tabular}{p{5cm}p{5cm}}                                                                                                                                                                                                                                                                                                                                                          \\
dependencies & version  \\
license & author/developer \\
maintainer & title/name \\
location/website/repo & language \\                                                                                                                                                                                                                                                                                                          
\end{tabular}
\end{table}

\begin{table}[]
\centering
\caption{Sample of unique metadata elements}
\label{tab:unique}
\begin{tabular}{p{5cm}p{5cm}}                                                                                                                                                                                                                                                                                                                                                          \\
preferred citation & bug tracker  \\
cost & category \\
cited by & keywords \\
ease of installation & unique identifier \\
inputs & number of collaborators \\                                                                                                                                                                                                                                                                                                                 
\end{tabular}
\end{table}

After each breakout group shared its results, there was general discussion amongst participants and a commitment from the organizers to make the results available, but otherwise, there was no firm conclusion nor action items identified.

\section{Conclusion}
This meeting provided some idea of the sort of additional data various people would like to see made available, and the discussion was interesting and useful. Some proposed metadata is straightforward, while some is dynamic (current support for a code, for example), and raises questions about the maintenance of metadata. Some, such as comments, represent opinions rather than data as such. It was clear that ASCL itself is not in the business of judging codes, in part because this would inhibit code release \citep{barnes2010}, though comments are of interest to users and those looking to build upon existing software.

\acknowledgements The ASCL is grateful for financial support provided by the Heidelberg Institute for Theoretical Studies (HITS) and for facilities and services support from Michigan Technological University, the Astronomy Department at the University of Maryland, and the University of Maryland Libraries.

\bibliography{B6}  

\end{document}